\documentclass[aps,prd,preprint,floatfix,nofootinbib,superscriptaddress]{revtex4}
\usepackage{graphicx,subfigure,epsfig,epsf}

\newcommand{\beq}{\begin{equation}}
\newcommand{\eeq}{\end{equation}}
\newcommand{\bea}{\begin{eqnarray}}
\newcommand{\eea}{\end{eqnarray}}
\newcommand{\bi}{\begin{itemize}}
\newcommand{\ei}{\end{itemize}}

\newcommand{\Lda}{\Lambda}
\newcommand{\g}{\gamma}
\def\Re{{\cal R \mskip-4mu \lower.1ex \hbox{\it e}\,}}
\def\Im{{\cal I \mskip-5mu \lower.1ex \hbox{\it m}\,}}

\def\etal{{\it et al.}}

\def\tev{\,{\ifmmode\mathrm {TeV}\else TeV\fi}}
\def\gev{\,{\ifmmode\mathrm {GeV}\else GeV\fi}}
\def\mev{\,{\ifmmode\mathrm {MeV}\else MeV\fi}}
\def\to{\rightarrow}

\begin{document}

\def\issue(#1,#2,#3){{\bf #1}, #2 (#3)} 

\def\APP(#1,#2,#3){Acta Phys.\ Polon.\ \issue(#1,#2,#3)}
\def\ARNPS(#1,#2,#3){Ann.\ Rev.\ Nucl.\ Part.\ Sci.\ \issue(#1,#2,#3)}
\def\CPC(#1,#2,#3){Comp.\ Phys.\ Comm.\ \issue(#1,#2,#3)}
\def\CIP(#1,#2,#3){Comput.\ Phys.\ \issue(#1,#2,#3)}
\def\EPJC(#1,#2,#3){Eur.\ Phys.\ J.\ C\ \issue(#1,#2,#3)}
\def\EPJD(#1,#2,#3){Eur.\ Phys.\ J. Direct\ C\ \issue(#1,#2,#3)}
\def\IEEETNS(#1,#2,#3){IEEE Trans.\ Nucl.\ Sci.\ \issue(#1,#2,#3)}
\def\IJMP(#1,#2,#3){Int.\ J.\ Mod.\ Phys. \issue(#1,#2,#3)}
\def\JHEP(#1,#2,#3){J.\ High Energy Physics \issue(#1,#2,#3)}
\def\JPG(#1,#2,#3){J.\ Phys.\ G \issue(#1,#2,#3)}
\def\MPL(#1,#2,#3){Mod.\ Phys.\ Lett.\ \issue(#1,#2,#3)}
\def\NP(#1,#2,#3){Nucl.\ Phys.\ \issue(#1,#2,#3)}
\def\NIM(#1,#2,#3){Nucl.\ Instrum.\ Meth.\ \issue(#1,#2,#3)}
\def\PL(#1,#2,#3){Phys.\ Lett.\ \issue(#1,#2,#3)}
\def\PRD(#1,#2,#3){Phys.\ Rev.\ D \issue(#1,#2,#3)}
\def\PRL(#1,#2,#3){Phys.\ Rev.\ Lett.\ \issue(#1,#2,#3)}
\def\PTP(#1,#2,#3){Progs.\ Theo.\ Phys. \ \issue(#1,#2,#3)}
\def\RMP(#1,#2,#3){Rev.\ Mod.\ Phys.\ \issue(#1,#2,#3)}
\def\SJNP(#1,#2,#3){Sov.\ J. Nucl.\ Phys.\ \issue(#1,#2,#3)}


\bibliographystyle{revtex}

\title{Laboratory frame analysis of $e^+ e^- \to \mu^+ \mu^-$ scattering in the Noncommutative Standard Model } 




\author{Prasanta~Kumar~Das}
\email[]{Author(corresponding) : pdas@bits-goa.ac.in}
\affiliation{Birla Institute of Technology and Science-Pilani, K. K. Birla Goa campus, NH-17B, Zuarinagar, Goa-403726, India }
\author{Abhishodh Prakash}
\email[]{Abhishodh.Prakash@stonybrook.edu}
\affiliation{Department of Physics and Astronomy, Stony Brook University, Stony Brook, New York 11790, USA}



\date{\today}

\begin{abstract}
We study the muon pair production $ e^+ e^- \to  \mu^+ \mu^-$ in the framework of the non-minimal noncommutative(NC) standard model to the second order of the NC parameter $\Theta_{\mu\nu}$. The $\mathcal{O}(\Theta^2)$ momentum dependent NC interaction significantly modifies the cross section and angular distributions which are different from the standard model. After including the effects of earth's rotation we analyse the time-averaged and time dependent observables in detail. The time-averaged azimuthal distribution of the cross section shows siginificant departure from the standard model. We find strong dependence of the total cross section(time- averaged) and their distributions on the orientation of the noncommutative electric vector (${\vec{\Theta}}_E$). The periodic variation of the total cross-section with time over a day seems to be startling and can be thoroughly probed at the upcoming Linear Collider(LC). 
\end{abstract}

\maketitle


\section{Introduction}
Several radical proposals based on the higher-dimensional spacetime(e.g. braneworld models) have been put forward that can narrow down the large hierarchy between the four-dimensional Planck scale $M_{Pl}$ and the electroweak scale $M_{EW}$. In the brane-world models where the gravity is strong at the TeV scale\cite{ADD98}, one can see the stringy 
effects and the effects of the space-time noncommutavity in the TeV colliders. Interests in the noncommutative(NC) field theory arose from the pioneering work of Snyder \cite{Snyder47} and the recent developments in string theories connected to the D-brane dynamics which at low-energy manifests that spacetime is noncommutative
\cite{Connes98,Douglas98,SW99}. 
Witten \etal \cite{Witten96} in 1996 has suggested that one can see some stringy effects by lowering the threshold value of commutativity to \tev, a scale which is not so far from present or future collider scale. 

~ The noncommutative spacetime can be characterized by the coordinate operators satisfying   
\beq \label{XXTheta}
[\hat{X}_\mu,\hat{X}_\nu]=i\Theta_{\mu\nu} =  \frac{i c_{\mu\nu}}{\Lda_{NC}^2}
\label{NCSTh}
\eeq
where, the matrix $\Theta_{\mu\nu}$ is real and antisymmetric. It  has the dimension of area and reflects the extent to which the spacetime is fuzzy i.e. noncommutative. In above $\Lda_{NC}~(\Lambda $, say) represents the NC scale and  $c_{\mu\nu}$ has the same properties as $\Theta_{\mu\nu}$.  In order to study an ordinary field theory in such a noncommutative space, one replaces the ordinary product of fields with Moyal-Weyl(MW) 
\cite{Douglas} $\star$ products defined by
\begin{equation}
(f\star
g)(x)=exp\left(\frac{1}{2}\Theta_{\mu\nu}\partial_{x^\mu}\partial_{y^\nu}\right)f(x)g(y)|_{y=x}.
\label{StarP}
\end{equation}
Using this we find the noncommutative quantum electrodynamics(NCQED) Lagrangian as
\begin{equation} \label{ncQED}
{\cal L}=\frac{1}{2}i(\bar{\psi}\star \gamma^\mu D_\mu\psi
-(D_\mu\bar{\psi})\star \gamma^\mu \psi)- m\bar{\psi}\star
\psi-\frac{1}{4}F_{\mu\nu}\star F^{\mu\nu} \label{NCL},
\end{equation}
where 
$D_\mu\psi=\partial_\mu\psi-ieA_\mu\star\psi$,$~~(D_\mu\bar{\psi})=\partial_\mu\bar{\psi}+ie\bar{\psi}\star
A_\mu$, and $F_{\mu\nu}=\partial_{\mu} A_{\nu}-\partial_{\nu}
A_{\mu}-ie(A_{\mu}\star A_{\nu}-A_{\nu}\star A_{\mu})$.  In the WM approach the group closure property is only found to hold for the $U(N)$ gauge theories and the matter content is found to be in the (anti)-fundamental and adjoint representations. In the Weyl-Moyal approach interestingly one finds 3-photons and 4-photons  vertices in NCQED analogous to the Yang-Mills theory. Using this method Hewett \etal \cite{Hewett01} explored several processes e.g. $e^+ e^- \to e^+ e^-$ (Bhabha), $e^- e^- \to e^- e^-$ (M\"{o}ller), $e^- \g \to e^- \g$, $e^+ e^- \to \g \g$ (pair annihilation), $\g \g \to e^+ e^-$ and $\g \g \to \g \g$ in the context of NCQED. Conroy \etal \cite{Hewett01} have investigated the process $e^+ e^- \to \gamma \to \mu^+ \mu^-$ in the context of NCQED and predicted a reach of $\Lda = 1.7$ TeV. In an effort to construct the noncommutative standard model(NCSM) one is restricted to $U(3)\otimes U(2) \otimes U(1)$ \cite{chaichian} and one requires a Higgs mechanism together with the introduction of additional gauge bosons in order to get the correct SM gauge group (after the removal of two $U(1)$ factors). An alternative is the minimal noncommutative standard model(mNCSM), in which the group closure property holds for the $SU(3)\otimes SU(2) \otimes U(1)$ gauge group and the corresponding Lie algebra becomes the enveloping algebra obtained via the Seiberg-Witten map (SWM). In the SWM \cite{Connes98, Douglas98,SW99,Witten96,Douglas,Jurco} approach the matter fields $\psi$ and the gauge field $A^\mu$ in the noncommutative spacetime can be expanded in terms of the commutative ones as power series in $\Theta$, 
\bea \label{SWM}
\hat{\psi}(x,\Theta) = \psi(x) + \Theta \psi^{(1)} + \Theta^2 \psi^{(2)} + \cdots \\
\hat{A_\mu}(x,\Theta) = A_\mu(x) + \Theta A_\mu^{(1)} + \Theta^2 A_\mu^{(2)} + \cdots 
\eea
The advantage in the SWM approach is that this construction can be applied to any gauge theory with arbitrary matter representation. Using the SW technique, Calmet \etal \cite{Calmet} first constructed a model (close to the standard model(SM)), with noncommutative gauge invariance which is known as the {\it minimal} noncommutative standard model(mNCSM) (as mentioned above). They listed the Feynman rules comprising the standard model interaction(modified) and new interactions which are absent in the SM. Intense phenomenological searches \cite{alboteanu} have been made to unravel several interesting features of this model.  Recently, one of us has investigated the impact of weak $Z$ and photon exchange in the Bhabha and the M\"{o}ller scattering in the noncommutative spacetime\cite{pdas}. However all these analyses are limited upto the first order in $\Theta$. It is necessary to go beyond $\mathcal{O}(\Theta)$. The present authors first reported a preliminary study in this direction: they analysed the $e^+ e^- \to \gamma,Z \to \mu^+ \mu^-$ to order $\Theta^2$ (without considering the effect of earth's rotation)\cite{abhishodh}.
  Now in a generic NCQED the triple photon vertex arises to order ${\mathcal{O}}(\Theta)$, which however is absent in the mNCSM. Another formulation of the NCSM came to the  forefront through the pioneering work by  Melic \etal \cite{Melic:2005ep} where the triple neutral gauge boson coupling \cite{Trampetic} appears naturally in the gauge sector. We will call this the nonminimal version of NCSM or simply NCSM. The Feynman rules to order $\mathcal{O}(\Theta)$ were presented in their work  \cite{Melic:2005ep}. In 2007, Alboteanu {\it et al} presented the $\mathcal{O}(\Theta^2)$ Feynman rules for the first time. In the present work we will confine ourselves within this nonminimal version of the NCSM and use the Feynman rules given in Alboteanu  \etal \cite{alboteanu}.
 
The noncommutativity parameter $\Theta_{\mu\nu}$ may be an elementary constant in nature that has a fixed direction in a specific coordinate system fixed to the celestial sphere. The laboratory frame which is located on the earth, is moving by earth's rotation. So we should take into account the apparent time variation of $\Theta_{\mu\nu}$ in the laboratory frame when we make any phenomenological investigation of scattering or decay of particles on the surface of the earth. In this paper, we consider the effect of earth's rotation on the muon pair production in the upcoming Linear Collider.   

 In addition if spacetime is anisotropic due to the noncommutativity, then a probe to the magnitude of the length scale (scale of anisotropy) and the specific direction of $\Theta_{\mu\nu}$ may be very interesting both from an experimental and theoretical point of view. We may determine the direction of $\Theta_{\mu\nu}$ by studying the behaviour of several time averaged observables. 

 In Sec~II, we describe the parametrization of the noncommutativity parameter $\Theta_{\mu\nu}$ including the effect of earth's rotation. We construct several time-averaged observables i.e. the cross section and the differential cross-section of$ e^+ e^- \stackrel{\gamma,Z}{\longrightarrow} \mu^+ \mu^-$. In Sec~III, we make a detailed numerical analysis and discuss the prospects of the TeV scale noncommutative geometry. Finally, in Sec~IV we summarize our results and conclude.   

\section{$ e^+ e^- \stackrel{\gamma,Z}{\longrightarrow} \mu^+ \mu^-$scattering in the laboratory frame}
\begin{figure}[htbp]
\vspace{5pt}
\centerline{\hspace{-3.3mm}
{\epsfxsize=14cm\epsfbox{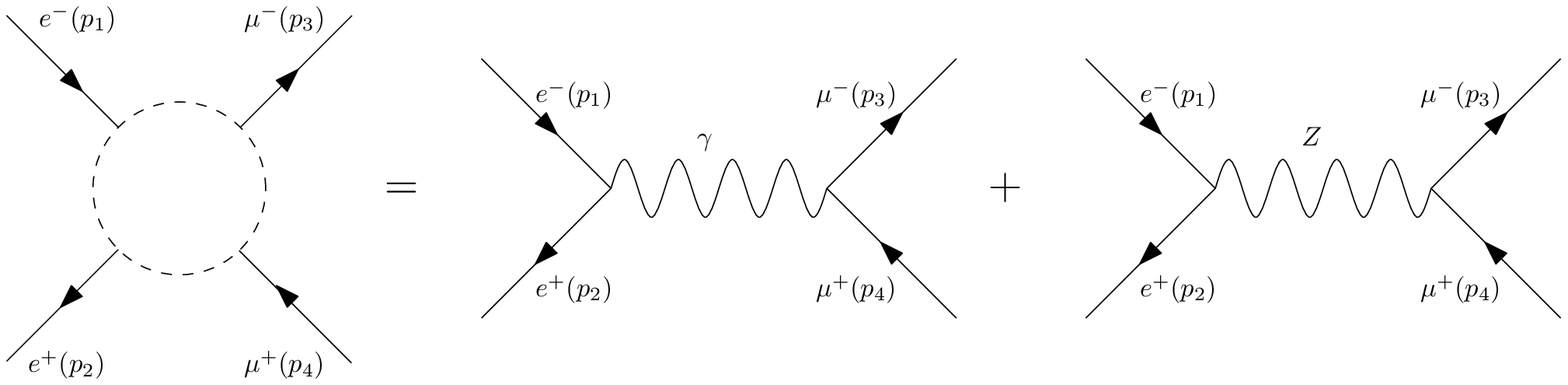}}}
\hspace{2.5cm}
\vspace*{-0.5in}
\caption{{\it Feynman diagrams for $ e^+ e^- \stackrel{\gamma,Z}{\longrightarrow} \mu^+ \mu^-$ in the NCSM.}}
\protect\label{feyn}
\end{figure}

\vspace{-0.15in}
The pair production of muons $e^- (p_1) e^+ (p_2) \stackrel{\gamma,Z}{\longrightarrow}  \mu^- (p_3) \mu^+ (p_4)$ occurs via the $s$ channel exchange of $\gamma$ and $Z$ bosons in the NCSM. The corresponding Feynman diagrams are shown below (Fig. \ref{feyn}). 
\noindent In order to have the cross section to order $\mathcal{O}(\Theta^2)$, we include the order $\mathcal{O}(\Theta^2)$ Feynman rule. The scattering amplitudes to order $\mathcal{O}(\Theta^2)$ for the photon mediated diagram can be written as  
\bea \label{gamma}
{\mathcal{A}}_\gamma = \frac{4 \pi \alpha}{s} \left[{\overline v}(p_2) \gamma_\mu u(p_1)\right]  
\left[{\overline u}(p_3) \gamma^\mu v(p_4)\right]  \times
 \left[(1 - \frac{(p_2 \Theta p_1)^2}{8}) + \frac{i}{2} (p_2 \Theta p_1)\right] \nonumber \\ \times \left[(1 - \frac{(p_4 \Theta p_3)^2}{8} ) + \frac{i}{2} (p_4 \Theta p_3)\right]
\eea
and for the $Z$ boson mediated diagram as 
\bea \label{Z}
{\mathcal{A}}_Z = \frac{\pi \alpha}{\sin^2(2\theta_W) s_Z}  \left[{\overline v}(p_2) \gamma_\mu (a + \gamma^5)  {\overline u}(p_1)\right] \times \left[{\overline u}(p_3) \gamma^\mu (a + \gamma^5)  {\overline v}(p_4)\right] \nonumber \\ \times  \left[(1 - \frac{(p_2 \Theta p_1)^2}{8}) + \frac{i}{2} (p_2 \Theta p_1)\right] \times \left[(1 - \frac{(p_4 \Theta p_3)^2}{8} ) + \frac{i}{2} (p_4 \Theta p_3)\right]
\eea

\noindent where $s=(p_1 + p_2)^2$, $\alpha = e^2/4\pi$, $a = 4 \sin^2(\theta_W) - 1$ and $\theta_W$ is the Weinberg angle. In the above
$s_Z = s - m_Z^2 - i m_Z \Gamma_Z$,  where $m_Z$ and $\Gamma_Z$ are the mass and decay width of the $Z$ boson. The Feynman rules required to evaluate such process are listed in Appendix A.

Since the noncommutative parameter $\Theta_{\mu\nu}$ is considered as fundamental constant in nature, it's direction is fixed with respect to an inertial(non rotating) coordinate system (which can be a celestial coordinate system). Now the experiment is done in the laboratory coordinate system which is located on the surface of the earth and is moving by the earth's rotation. This results in an aparent time variation of $\Theta_{\mu\nu}$ which should be taken into account before making any phenomenological investigations.  

 The effect of earth's rotation on noncommutative phenomenology were considered in several earlier studies \cite{NCearthrot}. Here we follow the work of Kamoshita \cite{NCearthrot}.
\begin{figure}[htbp]
\centerline{\hspace{-12.3mm}
{\epsfxsize=10cm\epsfbox{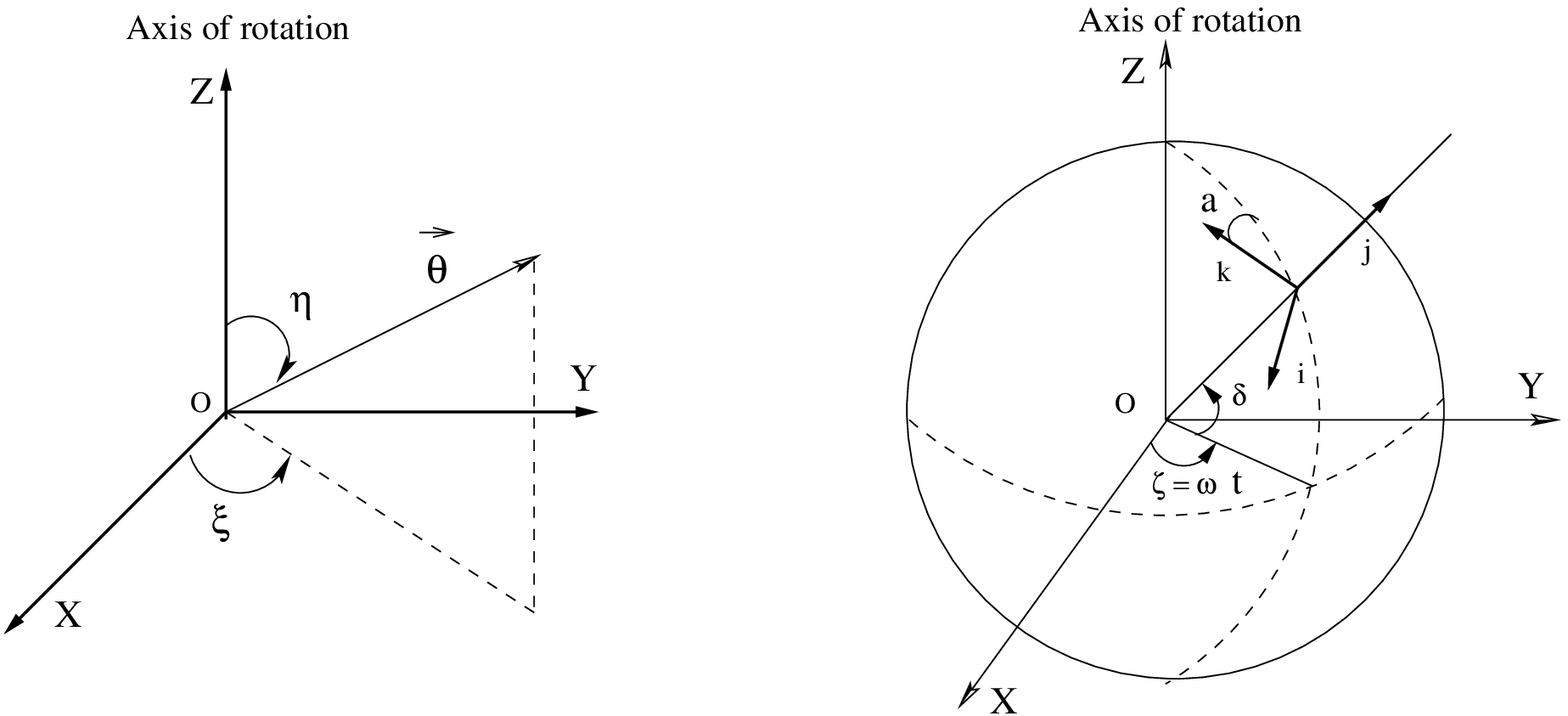}}}
\vspace*{-0.25in}
\caption{{\it In the left figure the primary coordinate system(X-Y-Z) is shown. The generic NC vector ${\vec{\Theta}}$ (electric or magnetic type)  of $\Theta_{\mu\nu}$ is shown with $\eta$ and 
$\xi$, respectively the polar and the azimuthal angle. In the right figure the arrangement of laboratory coordinate system (i-j-k) for an experiment on the earth in the primary coordinate system (X-Y-Z) is shown. In the above $\zeta = \omega t$ where $\omega$ is a constant. Also ($\delta$,~$a$), which defines the location of the laboratory, are constants.}}
\protect\label{sigplot}
\end{figure}
 Let ${\hat{i}}_X,~{\hat{j}}_Y$ and ${\hat{k}}_Z$ be the orthonormal basis of the primary(non rotating) coordinate system (X-Y-Z). Then in the laboratory coordinate system ($\hat{i} - \hat{j} - \hat{k}$), the bases vectors of the primary(non rotating) coordinate system can be written as  
\begin{eqnarray}
{\hat{i}}_X = \left(\begin{array}{c}
c_a s_\zeta + s_\delta s_a c_\zeta \\ c_\delta c_\zeta \\ s_a s_\zeta - s_\delta c_a c_\zeta
\end{array} \right), \nonumber 
{\hat{j}}_Y = \left(\begin{array}{c}
-c_a c_\zeta + s_\delta s_a s_\zeta \\ c_\delta s_\zeta \\ -s_a c_\zeta - s_\delta c_a s_\zeta
\end{array} \right), \nonumber 
{\hat{k}}_Z = \left(\begin{array}{c}
-c_\delta s_a  \\ s_\delta \\ c_\delta c_a 
\end{array} \right). \nonumber 
\end{eqnarray}
Here we have used the abbreviations $c_\beta = cos\beta,~s_\beta = sin\beta$ etc. In Fig. \ref{sigplot} we have shown the primary($X - Y - Z$) and laboratory($i - j - k$) coordinate system. 
Note that the primary $Z$ axis is along the axis of earth's rotation and ($\delta, a$) defines the location of $e^- - e^+$ experiment on the earth, with $- \pi/2 \le \delta \le \pi/2$ and $0 \le a \le 2 \pi$.  Because of earth's rotation the angle $\zeta$ (see Fig. 2) increases with time and the  detector comes to its original position after a cycle of one complete day, one can define $\zeta = \omega t$ with $\omega = 2 \pi/T_{day}$ and 
$T_{day} = 23h56m4.09053s$.
 Thus the electric and the magnetic components of the NC parameter $\Theta_{\mu\nu}$ in the primary system is given by 
\bea
{\vec{\Theta}}_E = \Theta_E (\sin\eta_E ~\cos\xi_E ~{\hat{i}}_X + \sin\eta_E ~\sin\xi_E ~{\hat{j}}_Y + \cos\eta_E ~{\hat{k}}_Z) \nonumber\\
{\vec{\Theta}}_B = \Theta_B (\sin\eta_B ~\cos\xi_B ~{\hat{i}}_X + \sin\eta_B ~\sin\xi_B ~{\hat{j}}_Y + \cos\eta_B ~{\hat{k}}_Z) \nonumber \\
\eea 
with
\beq
{\vec{\Theta}}_E = (\Theta^{01}, \Theta^{02}, \Theta^{03}), ~~ {\vec{\Theta}}_B = (\Theta^{23}, \Theta^{31}, \Theta^{12})
\eeq
and
\beq
\Theta_E = |{\vec{\Theta}}_E| = 1/\Lambda^2_E, ~~\Theta_B = |{\vec{\Theta}}_B| = 1/\Lambda^2_B.
\eeq
Here $(\eta, \xi)$ specifies the direction of the NC parameter $\Theta_{\mu\nu}$ w.r.t the primary coordinate system with $0 \le \eta \le \pi$ and $0 \le \xi \le 2 \pi$. In above $\Theta_E$ and $\Theta_B$ are the model parameters and the corresponding energy scale are defined by $\Lambda_E = 1/\sqrt{\Theta_E}$ and $\Lambda_B = 1/\sqrt{\Theta_B}$ which one can probe for different processes.

\noindent The spin-averaged squared-amplitude of the $ e^+ e^- \stackrel{\gamma,Z}{\longrightarrow} \mu^+ \mu^-$ scattering is given by
\beq \label{Ampsq}
{\overline {|{\mathcal{A}}|^2}} = {\overline {|{\mathcal{A}}_\gamma|^2}} + {\overline {|{\mathcal{A}}_Z|^2}} + 2 {\overline {Re({\mathcal{A}}_Z {\mathcal{A}}_\gamma^{ \dagger })}}. 
\eeq
The direct and interference terms in Eq.~\ref{Ampsq} are given in Appendix C. To calculate these we use the order ${\mathcal{O}}(\Theta^2)$ Feynman rules given in \cite{alboteanu}. It is interesting to note that all the lower order terms i.e. ${\mathcal{O}}(\Theta)$, ${\mathcal{O}}(\Theta^2)$, and ${\mathcal{O}}(\Theta^3)$  get cancelled (see the related discussion in the Appendix C and Appendix D of \cite{abhishodh})). Since it is difficult to get the time dependent data, we take the average of the cross section or it's distributionz over the sidereal day $T_{day}$ and compare that with the experimental data. We introduce the time averaged observables as follows:  
\bea \label{dsigma_avg}
\left<\frac{d^2\sigma}{d\cos\theta~d\phi}\right>_T &=& \frac{1}{T_{day}} \int_{0}^{T_{day}} \frac{d\sigma}{d\cos\theta~d\phi} dt, \\
\left<\frac{d\sigma}{d\cos\theta}\right>_T &=& \frac{1}{T_{day}} \int_{0}^{T_{day}} \frac{d\sigma}{d\cos\theta} dt, \\
\left<\frac{d\sigma}{d\phi}\right>_T &=& \frac{1}{T_{day}} \int_{0}^{T_{day}} \frac{d\sigma}{d\phi} dt, \\
\left<\sigma\right>_T &=& \frac{1}{T_{day}} \int_{0}^{T_{day}} \sigma dt,
\eea
where
\bea 
\label{sigma}
\sigma &=& \int_{-1}^1 d(\cos\theta) \int_0^{2 \pi} d\phi \frac{d \sigma}{d\cos\theta~d\phi}, \\
\label{dsdcostheta}
\frac{d\sigma}{d\cos\theta} &=& \int^{2 \pi}_0 d\phi \frac{d \sigma}{d\cos\theta~d\phi},  \\
\label{dsdphi}
\frac{d\sigma}{d\phi} &=& \int^1_{-1} d(\cos\theta) \frac{d \sigma}{d\cos\theta~d\phi}. 
\eea
In the above
\beq 
\frac{d^2 \sigma}{d\cos\theta~d\phi} = \frac{1}{64 \pi^2 s} {\overline {|{\mathcal{A}}|^2}}, 
\eeq
where $\sigma$ = $\sigma(\sqrt{s}, \Lambda, \theta, \phi, t)$. The time dependence in the cross section or it's distribution enters through the NC parameter ${\vec{\Theta}}(={\vec{\Theta}}_E)$ which changes with the change in $\zeta = \omega t $. The angle parameter $\xi$ appears in the expression of $\vec{\Theta}$ through $\cos(\omega t - \xi)$ or 
$\sin(\omega t - \xi)$ \cite{NCearthrot} as the initial phase for time evolution gets disappeared in the time averaged observables. So one can deduce ${\vec {\Theta}}_E $ i.e. $\Lambda_E$ and the angle $\eta_E$ from the time-averaged observables.    

\section{Numerical Analysis}
\vspace{-0.15in}
We describe in detail several characteristic results of the muon pair production in the NCSM and discuss at a length about how to probe the NC scale ${\vec{\Theta}}_E$ using cross-section and it's distributions in the laboratory coordinate system. Since it is difficult to get the time-dependent data, we consider the time-averaged total and differential cross-section and investigate their sensitivity on the NC scale $\Lambda~(=\Lambda_E)$ and the orientation angle $\eta~(=\eta_E)$). We set the laboratory coordinate system by taking $(\delta,a) = (\pi/4, \pi/4)$ which is the OPAL experiment at LEP. 
\subsection{Time-averaged angular distribution }
\vspace{-0.15in}
\noindent The angular distribution of the final state scattered particles is a useful tool to understand the nature of new physics. Since the noncommutativity of space-time defined by Eq. (\ref{XXTheta}) breaks Lorentz invariance including rotational invariance around the beam axis, this will lead to an anisotropy in the azimuthal distribution of the cross section i.e. the distribution will depend on $\phi$. This anisotropy which persists in the time-averaged (averaged over the side-real day $T_d$) azimuthal distribution of the cross section $\left<\frac{d\sigma}{d\phi}\right>_T$, can act as a clear signature of space time noncommutativity that is absent in the Standard Model and in many other extensions of it.   

\begin{figure}[htbp]
\centerline{\hspace{-12.3mm}
{\epsfxsize=7.0cm\epsfbox{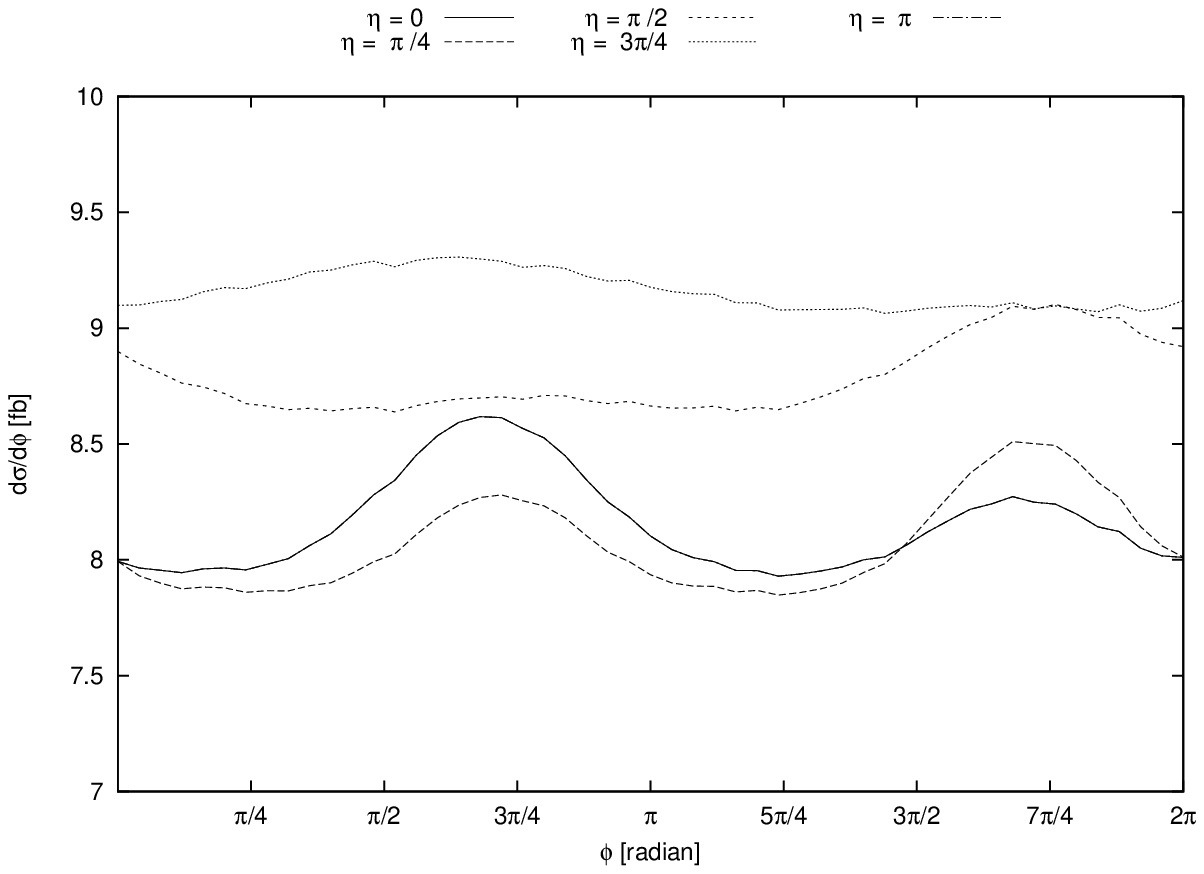}}  \hspace{2.0mm} {\epsfxsize=7.0cm\epsfbox{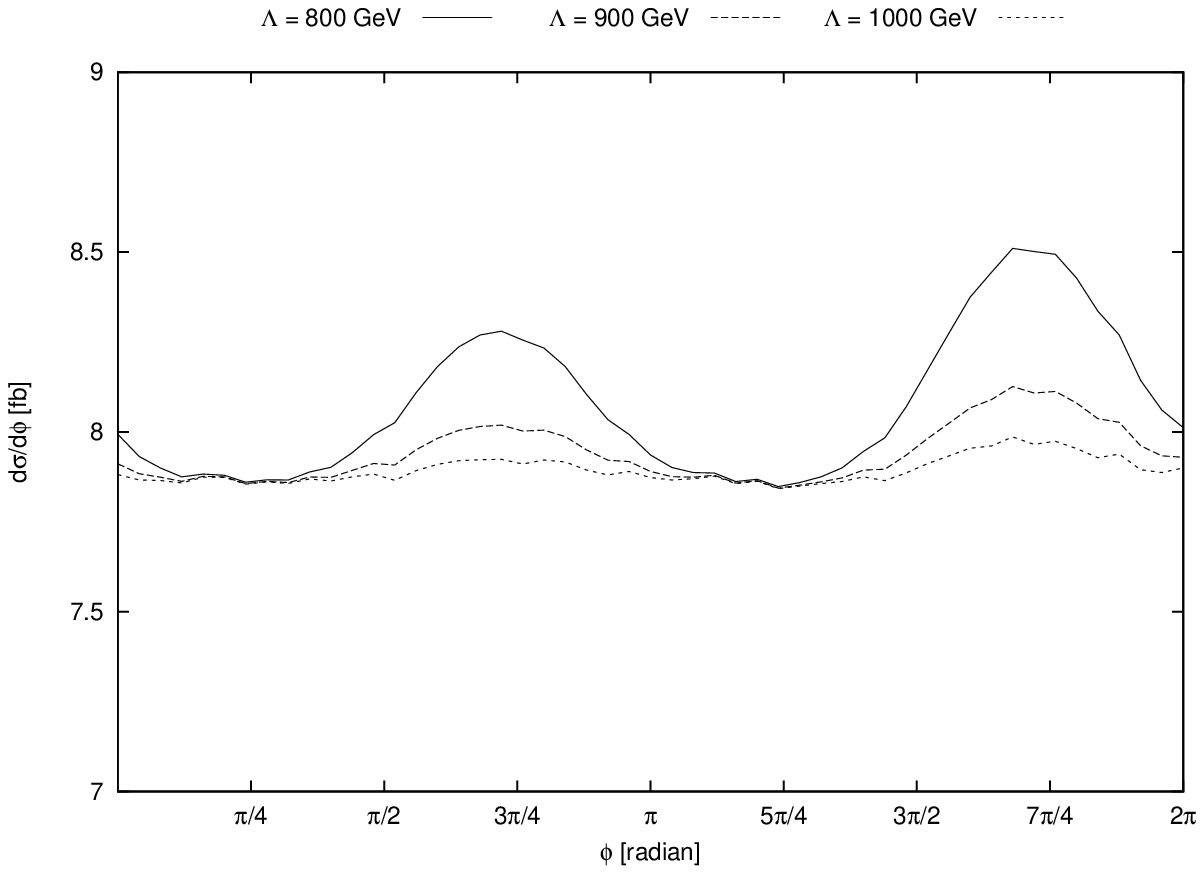}} } 
\vspace{-0.25in}
\caption{{{\it On the left $\left<\frac{d\sigma}{d\phi}\right>_T$ is plotted as a function of $\phi$ for $\eta=0,~\pi/4,~\pi/2,~3 \pi/4$ and $\pi$. We set the scale $\Lambda = 0.8$ TeV. On the right $\left<\frac{d\sigma}{d\phi}\right>_T$ is plotted as a function of $\phi$ corresponding to $\Lambda = 0.8,~0.9$ and $1.0$ TeV and $\eta = \pi/4$. For both plots we set the machine energy $E_{com} = 1.5$ TeV.}  }}
\protect\label{dsdphiplot_time}
\end{figure}

 In Fig. \ref{dsdphiplot_time} on the l.h.s we have plotted $\left<\frac{d\sigma}{d\phi}\right>_T$ vs the azimuthal angle $\phi$ for diferent values of 
$\eta$. We set the machine energy $\sqrt{s}(= E_{com}) = 1.5$ TeV and the NC scale $\Lambda$ at $0.8$ TeV. 
\vspace{-0.15in} 
The distribution has peaks corresponding to $\eta = 0, \pi/4, \pi/2$ and $\pi$. The plots on the l.h.s corresponding to $\eta = 0$ and $\pi$ are coincident. For $\eta = \pi/4$ the peak at the left (located at $\phi = 3 \pi/4$) is smaller than the peak at the right (located at $\phi = 7 \pi/4$) which is quite pronounced. For $\eta = 0 $ and $\pi$ it is opposite. The plot corresponding to $\eta =  \pi/2$ has a smaller peak at $\phi = 7 \pi/4$, whereas the plot corresponding to $\eta = 3 \pi/4$ is almost flat.
On the r.h.s of Fig. \ref{dsdphiplot_time}, we have plotted $\left<\frac{d\sigma}{d\phi}\right>_T$ as a function of $\phi$ with $\Lambda = 0.8,~0.9$ 
and $1.0$ TeV. We set $\eta = \pi/4$ and $E_{com} = 1.5$ TeV. Clearly the plot shows that there are two peaks: one at $\phi = 3 \pi/4$ and the other at 
$\phi = 7 \pi/4$.  The height of the peaks increases as one changes $\Lambda$ from $1.0$ TeV to $0.8$ TeV. Also the peaks on the right of a given plot are bigger than the ones on the left. This clearly shows the sensitivity of the distribution(signal) on the orientation ($\eta$) of the NC vector ${\vec{\Theta}}_E$, the orientation of the LC detector and the NC scale $\Lambda$.

\begin{figure}[htbp]
\centerline{\hspace{-12.3mm}
{\epsfxsize=7.0cm\epsfbox{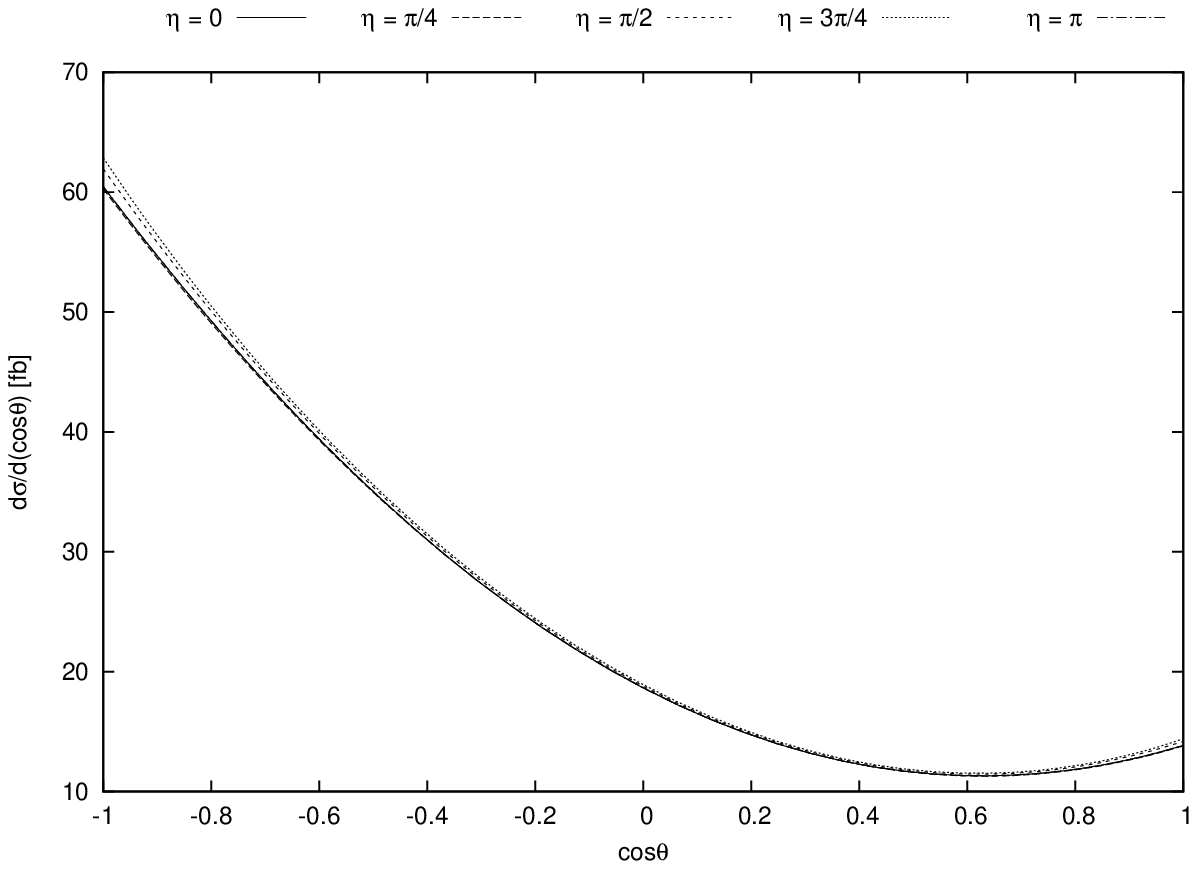}}  \hspace{2.0mm} {\epsfxsize=7.0cm\epsfbox{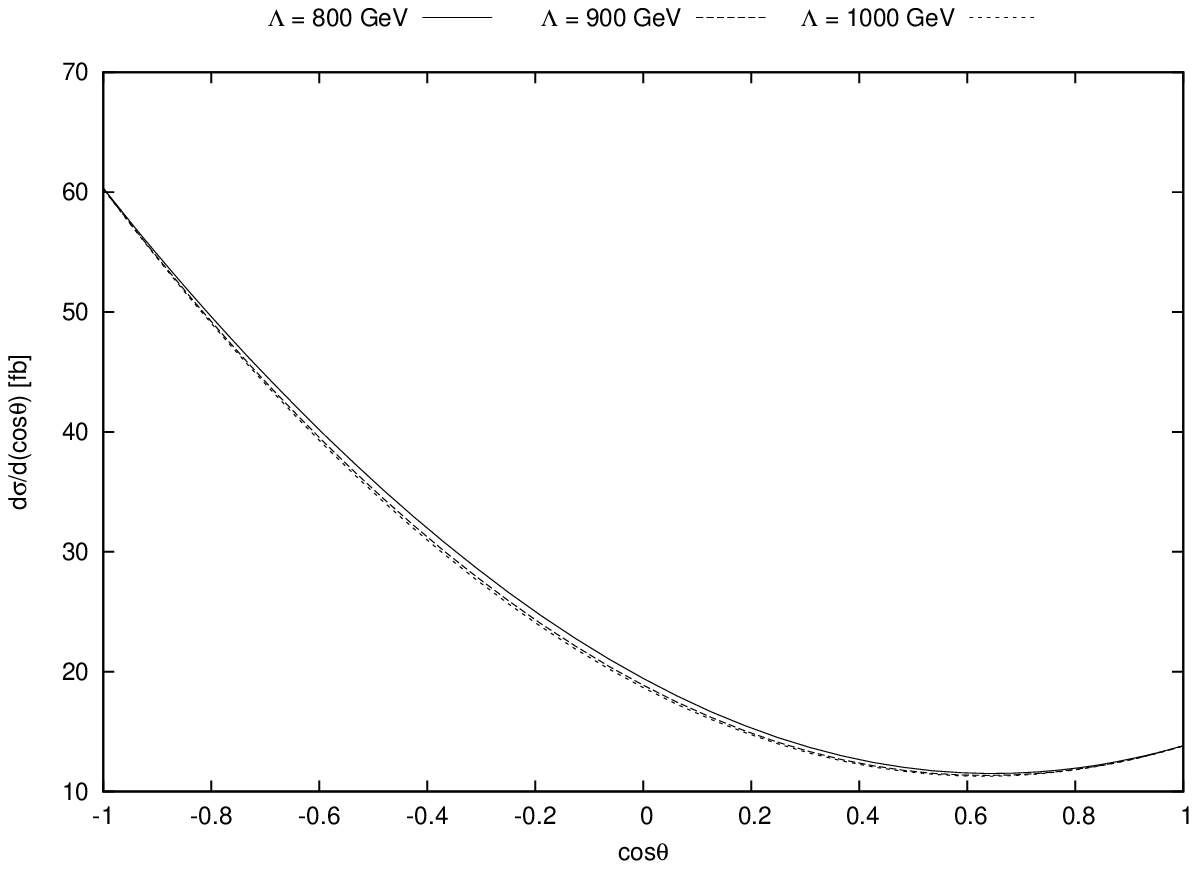}} } 
\vspace{-0.25in}
\caption{{{\it On the left $\left<\frac{d\sigma}{d\cos\theta}\right>_T$ is plotted against $\cos\theta$ corresponding to the NC scale $\Lambda = 0.8$ TeV and the orientation angle $\eta=0,~\pi/4,~\pi/2,~3 \pi/4$ and $\pi$. On the right $\left<\frac{d\sigma}{d\cos\theta}\right>_T$ is plotted against $\cos\theta$ for  $\Lambda = 0.8,~0.9$ and $1.0$ TeV and $\eta = \pi/4$. For both plots the machine energy is chosen to be $ E_{com} = 1.5$ TeV.}  }}
\protect\label{dsdcosthetaplot_time}
\end{figure}

 Next we consider the polar distribution $\left<\frac{d\sigma}{d\cos\theta}\right>_T$.  In Fig. \ref{dsdcosthetaplot_time} we have plotted $\left<\frac{d\sigma}{d\cos\theta}\right>_T$ as a function of $\cos\theta$, $\theta$ being the scattering angle of the final state muon($\mu^{-}$). On the l.h.s figure the plots correspond to $\Lambda = 0.8$ TeV and $\eta = 0,~\pi/4,\pi/2,~\pi/4$ and $\pi$, respectively. No significant changes in the polar distribution with $\eta$ is found.  On the r.h.s figure we have shown the plot of $\left<\frac{d\sigma}{d\cos\theta}\right>_T$ as a function of $\cos\theta$ with $\eta = \pi/4$ and $\Lambda = 0.8,~0.9$ and $1.0$ TeV.  So the time-averaged polar distribution is found to be insensitive to the orientation angle $\eta$ and the NC scale $\Lambda$. 
\vspace{-0.35in}
\subsection{Time averaged total cross section }
\begin{figure}[htbp]
\centerline{\hspace{-12.3mm}
{\epsfxsize=7.0cm\epsfbox{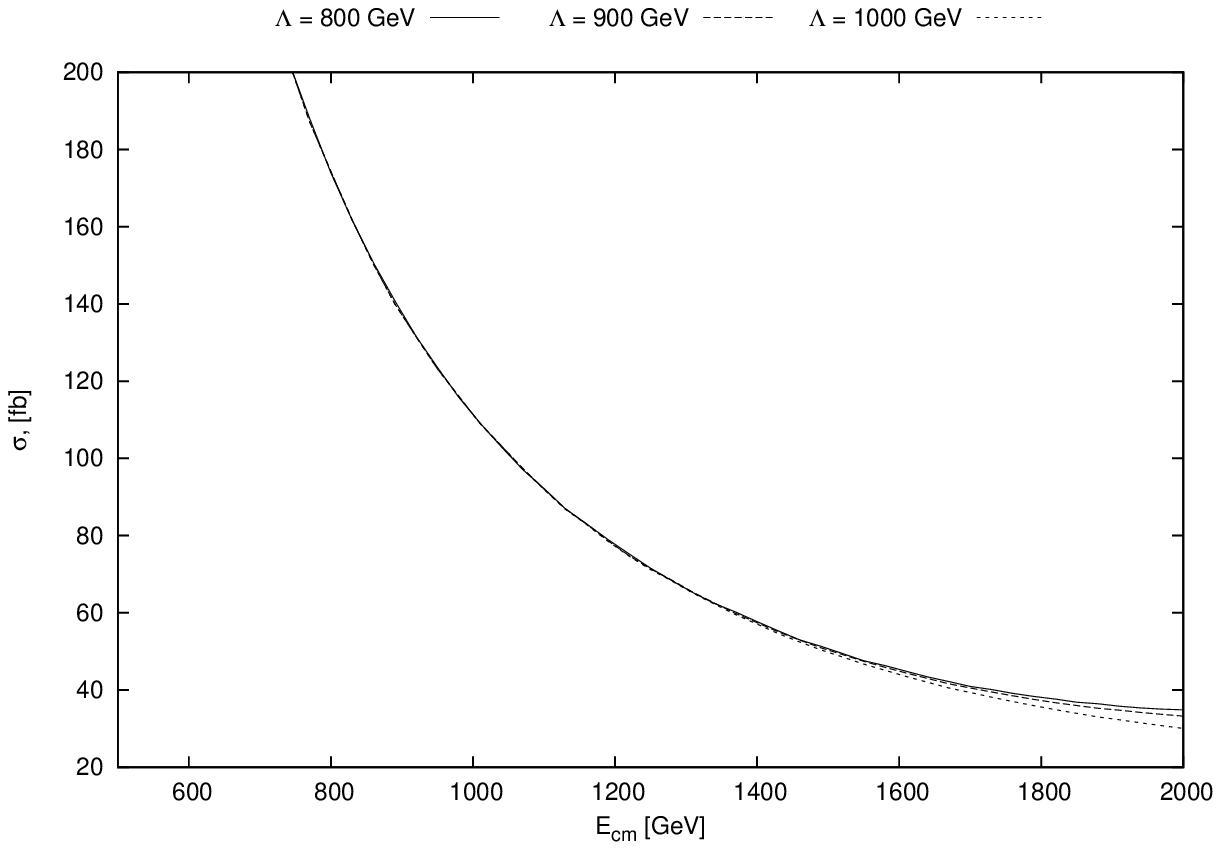}}  \hspace{2.0mm} {\epsfxsize=7.0cm\epsfbox{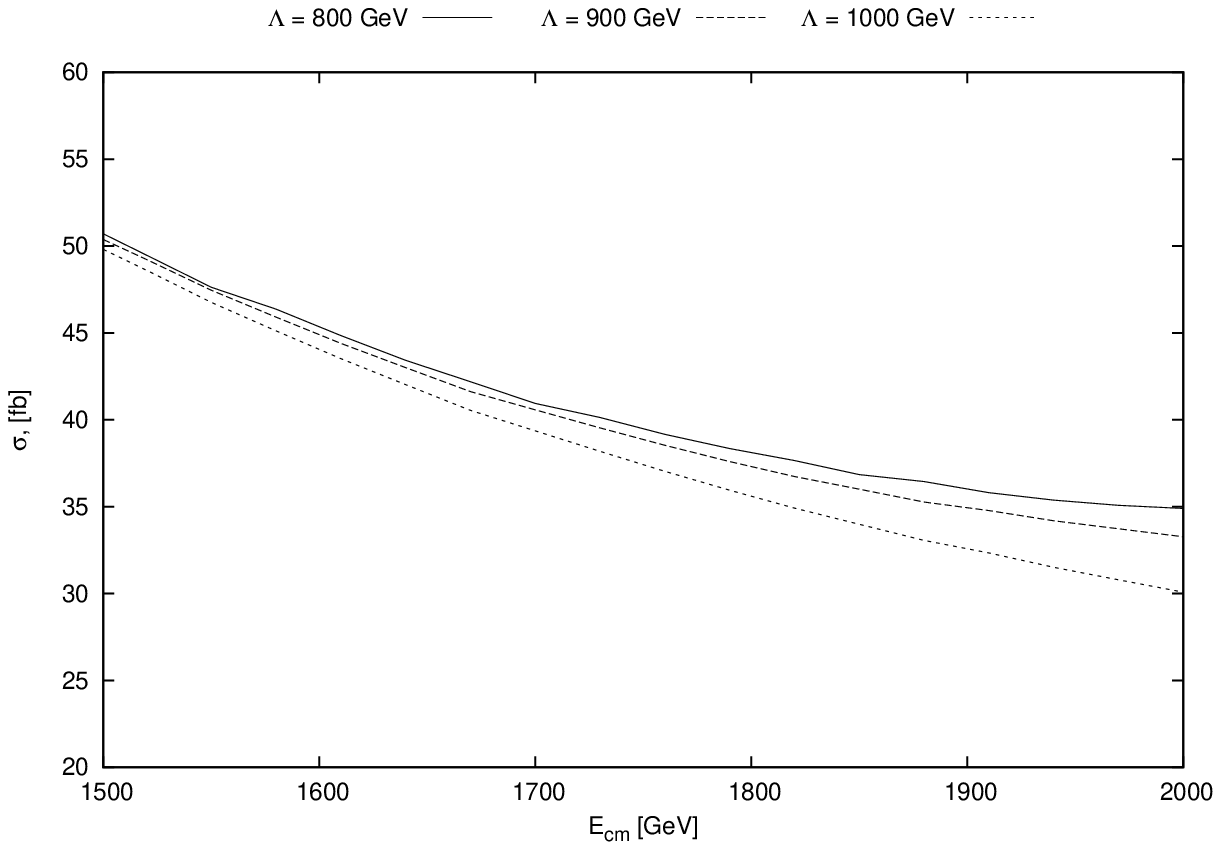}} } 
\vspace{-0.25in}
\caption{{{\it On the left the time-averaged total cross-section $\left<\sigma_T\right>_T$ ($fb$) is plotted as a function of $E_{com}$ corresponding to $\eta = \pi/4$ and $\Lambda = 0.8,~0.9$ and $1.0$ TeV. On the right the same quantity is plotted corresponding to $E_{com} = 1.5$ TeV to 
$E_{com} = 2.0$ TeV.}}}
\protect\label{sigmaenergyplot_time}
\end{figure}

\vspace{-0.15in}
In Fig. \ref{sigmaenergyplot_time} we have plotted the time-averaged total cross-section $\left<\sigma_T\right>_T$ (in $fb$) as a function of the machine energy $E_{com}$ (see the plots on the left) which varies from $0.5$ TeV to $2$ TeV. The three curves corresponding to $\Lambda = 0.8,~0.9,~1.0$ TeV and $\eta = \pi/4$ looks almost same: no significant deviation among them is found until the energy reaches $1.5$ TeV. On the right we have plotted the same with the machine energy($E_{com}$) varying from $1.5$ to $ E_{com} = 2.0$ TeV. We see a significant enhancement in the cross-section with the decrease in $\Lambda$.   
In Table 1, we make an estimate of the number of events per year (say, $N$) corresponding to different $\Lambda$. We set the machine energy $E_{com}$ at 
$1.5$ TeV and $2.0$ TeV with the integrated  luminosity ${\mathcal{L}} = 100~fb^{-1}$ for the LC.  We see that $N$ varies from  $5100~yr^{-1}$ to $5000 ~yr^{-1}$ corresponding to $E_{com} = 1.5$ TeV and from $3500~yr^{-1}$ to $3000 ~yr^{-1}$ for $E_{com} = 2.0$ TeV as $\Lambda$ changes from $0.8$ TeV to $1.0$ TeV.
\begin{center}
Table 1
\end{center}
\vspace*{-0.2in}
\begin{center}
\begin{tabular}{|c|c|c|c|}
\hline
$\Lambda$ (GeV) & NC signal ($\sigma$)(fb) & ${\mathcal{L}}(fb^{-1})$  & N (events per year)\\
\hline
\hline
  800  & 51 (35) & 100 & 5100 (3500) \\
\hline
  900 &  51 (33) & 100 & 5100 (3300) \\
\hline
  1000 & 50 (30) & 100 & 5000 (3000) \\
\hline
\end{tabular}
\end{center}
\noindent {\it Table 1: Progressive reduction of the NC signal and the number of events per year with the increase in $\Lambda$. We set the machine energy $E_{com} = 1.5$ TeV ($2.0$ TeV). We choose $\eta = \pi/4$ and the integrated luminosity of the LC about ${\mathcal{L}}=100~fb^{-1}$ $yr^{-1}$}.
\vspace{-0.25in}
\subsection{Time averaged total cross section as a function of $\eta$}
\begin{figure}[htbp]
\centerline{\hspace{-12.3mm}
{\epsfxsize=7.0cm\epsfbox{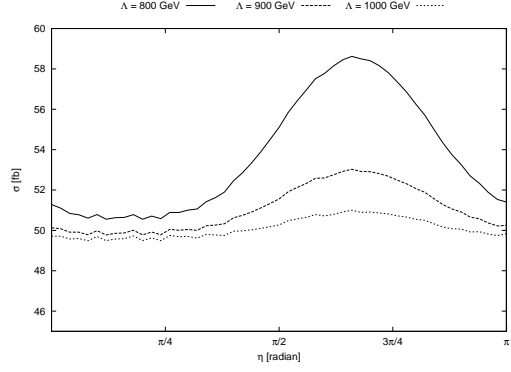}}  } 
\vspace{-0.25in}
\caption{{{\it The time-averaged cross-section $\left<\sigma_T\right>_T$ ($fb$) is plotted as a function of $\eta$ with the machine energy $E_{com} = 1.5$ TeV and the NC scale $\Lambda = 0.8,~0.9$ and $1.0$ TeV,respectively.  There are some characteristic peaks in the plot which are located at $\eta = 0.659 \pi$. }}}
\protect\label{sigmaetaplot}
\end{figure}

\vspace{-0.15in}
In Fig. \ref{sigmaetaplot} we have plotted $\left<\sigma_T\right>_T$ (time averaged total cross section) as a function of $\eta$ corresponding to $E_{com} = 1.5$ TeV and the NC scale $\Lambda = 0.8,~0.9$ and $1.0$ TeV. The plot shows that the cross-section has peaks at $\eta = 0.659 \pi$: the peak corresponding to $\Lambda = 0.8$ TeV is quite pronounced. The height of the peak decreases(and thus the NC signal) with the increase of $\Lambda$ from $0.8$ TeV to $1.0$ TeV
\vspace{-0.35in}
\subsection{Time varying total cross section }
\begin{figure}[htbp]
\centerline{\hspace{-12.3mm}
{\epsfxsize=7.0cm\epsfbox{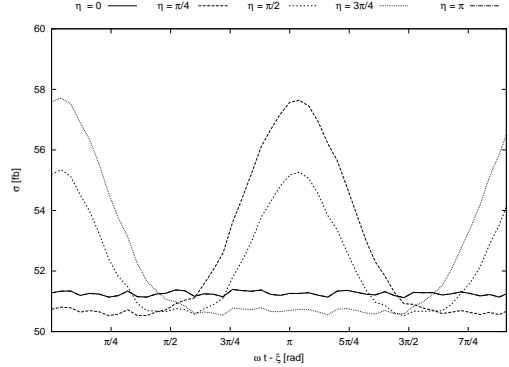}}  } 
\vspace{-0.25in}
\caption{{{\it The total cross section $\sigma$ is plotted as a function of 
$\zeta (= \omega t - \xi)$ (with $\xi = 0$) corresponding to $\eta = 0, ~\pi/4,~ \pi/2,~3 \pi/4$ and $\pi$.  We set $E_{com} = 1.5$ TeV and 
$\Lambda = 0.8$ TeV. There are peaks in the plot corresponding to $\eta = \pi/4$ and $\pi/2$ which are located at $\zeta = 3.27$ rad.} }}
\protect\label{sigmatimeplot}
\end{figure}

\vspace{-0.15in}
Finally we look at the time dependent behaviour of the cross-section. In Fig. \ref{sigmatimeplot} we have plotted $\sigma$(unaveraged) as a function of $\zeta (= \omega t - \xi)$ corresponding to $\eta = 0, ~\pi/4, \pi/2,~3 \pi/4$ and $\pi$.
Here $\omega = 2 \pi/T_{day}$ with $T_{day} = 23h56m4.09053s$. We set the machine energy $E_{com} = 1.5$ TeV, $\Lambda = 0.8$ TeV and $\xi = 0$ in $\zeta$.
We make the following observations:
\bi
\item The cross section $\sigma$ is found to be completely flat for $\eta = 0$ and $\pi$: no change in the cross section as a function of $\zeta$ over the period of the complete day $T_{day}$. 
\item The cross-section  $\sigma$ has a pronouned peak at $\omega t = 3.27$ rad corresponding to $\eta = \pi/4$. Beyond this particular $\zeta$ it falls everywhere. Similarly, the cross-section corresponding to $\eta = 3 \pi/4$  shows exactly the opposite behaviour obtained for $\eta = \pi/4$. At $\omega t = 3.27$ rad it is found to be flat, whereas on either side it increases and possesses maximum at $\omega t = 0.12$~rad and $6.28$~rad, respectively.
\item Finally we see that the unaveraged cross section $\sigma$(corresponding to $\eta = \pi/2$) have alternate maxima and minima which at $t = 0.12/\omega,~3.27/\omega$ and $6.28/\omega$ sec i.e. different times of a day. This periodic cross-section (and thus the signal) is quite interesting and the finding of such event at the Linear Collider will validate the idea of the space-time noncommutativity at the TeV scale.  
\ei
\section{Conclusion}
\vspace{-0.15in}
We have investigated the effect of space-time noncommutativity on the fundamental processes  $ e^+ e^- \stackrel{\gamma,Z}{\longrightarrow} \mu^+ \mu^-$ with ${\mathcal{O}}(\Theta^2)$ Feynman rules taking the earth rotation into account. Since it is difficult to get time-dependent data, we have constructed several time-averaged observables and investiagated their sensitivities on the NC scale $\Lambda (=\Lambda_E)$ and the orientation angle $\eta(=\eta_E)$ of the eletric NC vector ${\vec{\Theta}}_E$. The pronounced peaks observed in the time-averaged azimuthal distribution are located at $\phi = 3 \pi/4$ and $7 \pi/4$ corresponding to the machine energy $E_{com} = 1.5$ TeV, $\Lambda = 0.8$ TeV and $\eta = 0,~\pi/4,~\pi/2$ and $\pi$. The peaks, not to be found in the standard model and in many of it's extensions , can act as a clear signature of space-time noncommutativity. In the azimuthal distribution the strength of the NC signal gets enhanced with the decrease in $\Lambda$.  The polar distribution (time-averaged) is found to be insensitive to $\Lambda$ and $\eta$. The asymmetry around the $\cos\theta = 0$ persists even when the space-time is noncommutative.  Assuming the integrated luminosity of the Linear Collider about $100~fb^{-1}$, we find that the number of events varies from $5100~yr^{-1}$ to $5000~yr^{-1}$ as $\Lambda$ changes from $1$ TeV to $0.8$ TeV at a machine energy $E_{com} = 1.5$ TeV. We have shown the dependence of the time-averaged total cross-section $\left<\sigma_T\right>_T$ on the orientation of the NC vector corresponding to the machine energy $E_{com} = 1.5$ TeV and $\Lambda = 0.8,~0.9$ and $1.0$ TeV, respectively. The plot shows  pronounced peak at $\eta = 0.659~\pi$ that do reflect the sensitivity of $\left<\sigma_T\right>_T$ on $\eta$. Finally, we have shown how the total cross-section $\sigma$(unaveraged) varies with $\zeta (=\omega t - \xi)$ (with $\xi = 0$). At certain time of a day the cross-section(unaveraged) and hence the NC signal becomes maximum. Correponding to $\eta = \pi/4$, we find that the cross-section is maximum at $\omega t =  3.27$ rad, whereas for $\eta = 3 \pi/4$ it aprrears at $\omega t = 0.12$~rad and $6.28$~rad, respectively. The cross-section $\sigma$ corresponding to $\eta = \pi/2$ is found to posses alternate maxima and minima which appears during different times of a day. The periodic appearence of maxima and minima of the NC signal is a novel prediction which can be tested in the upcoming Linear Collider. 
 Thus the noncommutative geometry is quite rich in terms of its phenomenological implications and there are a lot of potentially relevant interesting processes which can be probed at the Linear Collider. 
\vspace*{-0.15in}
\begin{acknowledgments}
The work of P.K.Das is supported by the SEED Project 2011 and DST YS Project SR/FTP/PS-11/2006. The authors would like to acknowledge Mr. Anupam Mitra who was involved in the initial part of this project. P.K.Das would to thank Santosh Kumar Rai of Oklahoma University,USA for the useful comments. 
\end{acknowledgments}
\vspace*{-0.15in}
\appendix

\section{Feynman rules to order ${\mathcal{O}}(\Theta^2)$ }
\noindent We follow the reference \cite{alboteanu} for ${\mathcal{O}}(\Theta^2)$ Feynman rules. The Feynman rule for the vertex for $\gamma(k) \to f (p_{in}) f(p_{out})$ vertex (where $f$ stands for fermion) can be written as $ i g V_\mu(p_{out}, k , p_{in})$, where
\bea
V_\mu = V_\mu^0 + V_\mu^1 + V_\mu^2
\eea
where, $V_\mu^0$ corresponds to the standard model vertex factor and $V_\mu^1$,$V_\mu^2$ correspond to corrections to orders $\Theta$ and $\Theta^2$, respectively. Using the above mentioned reference, we can evaluate the correction terms (setting  $c_A^{(1)}=c_\psi^{(1)}=0 $ and $c_\psi^{(2)}=0$ and using the momentum conservation rule $p_{in} + k = p_{out}$:
\bea
V_\mu^{(1)} = \frac{i}{2} [(p_{out}\Theta)_\mu \not{p_{in}}  + (\Theta p_{in})_\mu \not{p_{out}} - (p_{out} \Theta p_{in})\gamma_\mu) ]. \\
V_\mu^{(2)} = \frac{1}{8} (p_{out} \Theta p_{in}) [(p_{out}\Theta)_\mu \not{p_{in}}  + (\Theta p_{in})_\mu \not{p_{out}} - (p_{out} \Theta p_{in})\gamma_\mu) ].
\eea
With these, we can obtain  the Feynman rule for $\gamma$ and $Z$ exchanges as:
\bea
\gamma: i e Q_f\left[ \gamma_\mu + \left( \frac{i}{2} + \frac{(p_{out} \Theta p_{in})}{8}\right)\left[(p_{out}\Theta)_\mu \not{p_{in}}  + (\Theta p_{in})_\mu \not{p_{out}} - (p_{out} \Theta p_{in})\gamma_\mu) \right]\right] \nonumber \\ \\
Z: \frac{i e}{sin2\theta_W}  \left[ \gamma_\mu \Gamma^-_A + \left( \frac{i}{2} + \frac{(p_{out} \Theta p_{in})}{8}\right)\left[(p_{out}\Theta)_\mu \not{p_{in}}\Gamma^-_A   + (\Theta p_{in})_\mu \not{p_{out}}\Gamma^-_A  - (p_{out} \Theta p_{in})\gamma_\mu \Gamma^-_A ) \right]\right], \nonumber \\ 
\eea
\noindent where $\Gamma^-_A = c_V - c_A \gamma^5$ with $c_V = T_3 - 2 Q_f s_w^2$ and $c_A = T_3$.
The $\Theta$ weighted momentum dot product $ p_{out} \Theta p_{in} = p_{out}^\mu  \Theta_{\mu \nu}  p_{in}^\nu = -p_{in} \Theta p_{out}$.

\section{Momentum prescriptions and dot products}
We work in the center of momentum frame where the 4 momenta of the incoming and outgoing particles are given by:
\bea
\label{prescstart} p_1 &=& p_{e^-} = \frac{\sqrt{s}}{2}\left(1, 0, 0, 1\right)\\
p_2 &=& p_{e^+} = \frac{\sqrt{s}}{2} \left(1, 0, 0, -1\right)\\
p_3 &=& p_{\mu^-} = \frac{\sqrt{s}}{2}\left(1,\sin\theta \cos\phi,\sin\theta \sin\phi, \cos\theta \right)   \\
p_4 &=& p_{\mu^+} = \frac{\sqrt{s}}{2}\left(1,-\sin\theta \cos\phi,- \sin\theta \sin\phi, - \cos\theta \right),
\eea
where $\theta$ is the scattering angle made by the $3$-momentum vector $p_3$ of $\mu^-(p_3)$ with the $\hat{k}$ axis (the $3$-momentum direction of the incoming electron $e^-$) and $\phi$ is the azimuthal angle. 

 The antisymmetric NC tensor $\Theta_{\mu \nu} = ({\vec{\Theta}}_E,~{\vec{\Theta}}_B)$ i.e. it has $3$  electric and $3$ magnetic components. The $s$-channel driven muon pair production in electron-positron collision is found to be sensitive only to the ${\vec{\Theta}}_E$ vector and hence one obtain constraints on $\Lambda_E$ ($\Lambda$). 
In the laboratory frame (with $\eta = \eta_E$,~$\xi = \xi_E$ ), the electric NC vector ${\vec{\Theta}}_E $ can be written as 
\bea
{\vec{\Theta}}_E  &=& \Theta_E \sin\eta ~\cos\xi ~{\hat{i}}_X + \Theta_E \sin\eta ~\sin\xi ~{\hat{j}}_Y + \Theta_E \cos\eta ~{\hat{k}}_Z \nonumber \\
                 &=& \Theta^{lab}_{Ex} {\hat{i}} + \Theta^{lab}_{Ey} {\hat{j}} + \Theta^{lab}_{Ez} {\hat{k}}
\eea 
where
\bea
\Theta^{lab}_{Ex} &=& \Theta_E \left( s_\eta c_\xi (c_a s_\zeta + s_\delta s_a c_\zeta) + s_\eta s_\xi (-c_a c_\zeta + s_\delta s_a s_\zeta) - c_\eta c_\delta s_a)  \right) \nonumber \\
\Theta^{lab}_{Ey} &=& \Theta_E \left( s_\eta c_\xi c_\delta c_\zeta + s_\eta s_\xi c_\delta s_\zeta + c_\eta s_\delta \right) \nonumber \\
\Theta^{lab}_{Ez} &=&  \Theta_E \left( s_\eta c_\xi (s_a s_\zeta - s_\delta c_a c_\zeta) - s_\eta s_\xi (s_a c_\zeta + s_\delta c_a s_\zeta) + c_\eta c_\delta c_a)  \right)
\eea
with
\beq
\Theta_E = |{\vec{\Theta}}_E| = 1/\Lambda^2.
\eeq
In the above we have used abbreviations viz $s_\eta = \sin\eta,~c_\xi = \cos\xi$ etc.  As mentioned before, $(\eta, \xi)$ specifies the direction of ${\vec{\Theta}}_E$ w.r.t the primary coordinate system with $0 \le \eta \le \pi$ and $0 \le \xi \le 2 \pi$.  Using these we find
\bea
p_2 \Theta p_1 &=&  -\frac{s}{2} \Theta^{lab}_{Ez}, \\
p_4 \Theta p_3 &=&  -\frac{s}{2} \left(s_\theta c_\phi \Theta^{lab}_{Ex} + s_\theta s_\phi \Theta^{lab}_{Ey} + c_\theta \Theta^{lab}_{Ez} \right).
\eea
where $\Theta^{lab}_{Ex},~\Theta^{lab}_{Ey}$ and $\Theta^{lab}_{Ez}$ are defined above.
\section{Spin-averaged squared-amplitude for $ e^+ e^- \stackrel{\gamma,Z}{\longrightarrow} \mu^+ \mu^-$}
\noindent The squared-amplitude terms of Eq.\ref{Ampsq} are
\beq
\overline {|{\mathcal{A}}|^2} = {\overline {|{\mathcal{A}}_\gamma|^2}} + {\overline {|{\mathcal{A}}_Z|^2}} + 2 {\overline {Re({\mathcal{A}}_Z {\mathcal{A}}^{\dagger}_{\gamma})}} = \frac{1}{4} |{\mathcal{A}}|^2 . 
\eeq


\noindent where the direct and the interference terms are given by 
\bea
{\overline {|{\mathcal{A}}_\gamma|^2}} &=& 16 \pi^2 \alpha^2 A_{NC} (1 + \cos^2\theta),  \\
{\overline{|{\mathcal{A}}_Z|^2}} &=& \frac{4 \pi^2 \alpha^2}{\sin^4(2\theta_W)} \frac{s^2~A_{NC}}{[(s-m_Z^2)+\Gamma_Z^2m_Z^2]} [(1 - 4 s^2_W + 8 s^4_W)^2 (1 + \cos^2\theta) + 2 (1 - 4 s^2_W)^2 \cos\theta],  \nonumber \\ \\
2 Re (\overline{A_Z A_\gamma^\dagger}) &=& \frac{8 \pi^2 \alpha^2}{\sin^2 (2\theta_W)} \frac{s(s-m_Z^2)~A_{NC}}{[(s-m_Z^2)^2 + \Gamma_Z^2 m_Z^2]} [(1 - 4 s^2_W)^2 (1 + \cos^2\theta) + 2 \cos\theta ], \nonumber \\ 
\eea
where $s_W = \sin \theta_W$ and 
\bea
A_{NC}  
= \left[1 + \frac{(p_2 \Theta p_1)^4}{64} \right] \left[1 + \frac{(p_4 \Theta p_3)^4}{64} \right]
\eea 

\end{document}